\documentclass[12pt]{article}
\usepackage{amsmath,amssymb,theorem,cite,epsfig,url,psfrag,eepic,amsmath,mathtools,mathrsfs,amsbsy,dsfont,esint,braket,cancel}

\usepackage{indentfirst}
\usepackage{graphicx}
\usepackage{tikz,pgfplots}
\usepackage{bm,upgreek}
\usepackage[bookmarks=true,hyperfigures=true,colorlinks=true,linkcolor=black,citecolor=black,urlcolor=black,bookmarksnumbered,hidelinks]{hyperref}
\def\XXint#1#2#3{{\setbox0=\hbox{$#1{#2#3}{\int}$}
     \vcenter{\hbox{$#2#3$}}\kern-.5\wd0}}

\advance \topmargin by -\headheight
\advance \topmargin by -\headsep     
\evensidemargin \oddsidemargin
\def\Maketitle{{\def\newpage{}\maketitle}}
\begin{document}
%\rightline{\texttt{\today}}
\title{\textbf{{Correlation functions of degenerate fields in Super-Liouville field theory}
}\vspace*{.3cm}}
\date{}
\author{Aleksandra Ivanova$^{1}$ 
\\[\medskipamount]
\parbox[t]{0.85\textwidth}{\normalsize\it\centerline{1. HSE University, 6 Usacheva str., Moscow 119048, Russia}}
%\\
%\parbox[t]{0.85\textwidth}{\normalsize\it\centerline{2. Krichever Center, Skolkovo Institute of Science and Technology, 121205 Moscow, Russia}}
%\\
%\parbox[t]{0.85\textwidth}{\normalsize\it\centerline{3. Landau Institute for Theoretical Physics, 142432 Chernogolovka, Russia}}
}
\Maketitle
\begin{abstract}
We study four-point correlation functions of degenerated fields in the $NS$ sector in Super-Liouville field theory. We find integral expressions for these functions using the BPZ equation, and study some superconformal properties of these solutions. Finally, we present the general form for three-point correlation functions.
\end{abstract}
%%%%%%%%%%%%%%%%%%%%%%%%%%%%%%%%%%%%%%%%%%%%%%%%%%%%%%%%%%%%%%%%%%%%%%%%%%%%%%%%%%%%%%%%%%%%%%%%%
%\tableofcontents
\section{Introduction}
Liouville field theory (LFT) plays an important role in the description of non-critical string theories \cite{polyakov1981quantumb,polyakov1981quantumf}. At the same time, supersymmetric extensions are of particular interest. One of the most important steps is to study and generalize the methods used in bosonic theories.

One of the main problems of any quantum field theory is the calculation of correlation functions. As it has been shown in \cite{goulian1991correlation}, correlation functions in LFT can be calculated using a method very similar to the representation of the Coulomb gas \cite{dotsenko1985four}.

There is another method for calculating correlation functions in Conformal Field Theory (CFT). This method is the following. It is well known that correlation functions containing at least one degenerate field obey some linear differential equations known as BPZ equations \cite{belavin1984infinite}. In many special cases, when degeneration occurs at low levels, four-point correlation functions can be obtained by directly solving these differential equations, thus avoiding Coulomb gas type representations. This is important because the charge balance condition for Coulomb gas affects all fields included in correlation function contrary to the degeneracy condition, that applies only to a special field. By exploring the crossing symmetry of those correlation functions, a set of functional relations on generic three point functions can be derived. This method is known as Teshner's trick \cite{teschner1995liouville}. In \cite{poghossian1997structure}, the Teschner's trick was used in Super-Liouville field theory for a degenerate vector on the second level in the Ramond sector. In our work, this method is applied to a vector at the $3/2$ level in Neveu–Schwarz sector.

In this paper, we consider four-point correlation function in the $NS$ sector of the $N = 1$ supersymmetric Liouville field theory. This paper is organized as follows. In Section 2 we give a brief overview of $N=1$ supersymmetric CFT. Section 3 considers the subsystem of four-point correlation functions with an even number of fermions. Explicit expressions for all correlation functions in terms of the four-point bosonic function are given. After this, a similar analysis is performed for the odd subsystem. Section 4 shows that the selected bosonic function involving a degenerate field at the $3/2$ level corresponds to a specific differential equation. At the same time, the solution of this equation has an integral form. Section 5 considers the superconformal properties of operator product expansions (OPE). Correlation functions are represented through structure constants related to the field fusion rules and to three-point functions. Ratios of these constants are found. Finally, Section 6 presents the general form of three-point correlation functions that received with help of functional equations for constants of three-point functions. 

\section{Super-Liouville field theory}
Super-Liouville field theory (SLFT) is a supersymmetric generalization of the bosonic Liouville theory, which is known to be a theory of gravity in two dimensions. Similarly, SLFT describes 2d supergravity caused by supersymmetric matter. Let us give some information about this theory (more details can be found in \cite{friedan1985superconformal,mussardo1988fine,fukuda2002super}).

To obtain the SLFT action, one can simply ``supersymmetrize" the bosonic Liouville action \cite{polyakov1981quantumf,fukuda2002super,distler1990super}. The action takes the following form:
\begin{align} \label{(2.1)}
    S = \frac{1}{2 \pi} \int d^2 z \bigg( \partial \phi \bar{\partial} \phi + \bar{\psi} \partial \bar{\psi} + \psi \bar{\partial} \psi \bigg) + 2i \mu b^2 \int d^2 z \psi \bar{\psi} e^{b \phi} , 
\end{align}
where $b$  is the coupling constant related to the background charge $Q$ and the central charge $\hat{c}$ in the following way\footnote{Here $\hat{c} = \frac{2}{3} c$, where c is the conventional central charge.}
\begin{align*}
    Q &= b + \frac{1}{b}, \hspace{20pt} \hat{c}= 1 + 2 Q^2 .
\end{align*}

The infinite superconformal algebra in 2D is generated by the energy-momentum tensor $T(z)$ and the supercurrent $G(z)$. The singular part of the operator product expansion (OPE) of $T$ and $G$ corresponds to the $N = 1$ superconformal algebra
\begin{align}
    &T(z)T(w) = \frac{3 \hat{c}}{4 (z-w)^4} + \frac{2 T(w)}{(z-w)^2} + \frac{T'(w)}{z-w} + ... ,  \nonumber\\
    &T(z)G(w) = \frac{3 G(w)}{2 (z-w)^2} + \frac{G'(w)}{z-w} + ... , \\
    &G(z)G(w) = \frac{\hat{c}}{(z-w)^3} + \frac{2 T(w)}{z-w} + ...  \nonumber 
\end{align}
The relations above are written for the holomorphic part, similar expressions are also valid for the antiholomorphic one. The following identities will be given only for the holomorphic part. The primary fields are defined as
\begin{align}
    &T(z) \Phi(w) = \frac{\Delta \Phi(w)}{(z-w)^2} + \frac{\partial \Phi(w)}{z-w} + ... , \\
    &T(z) \Psi(w) = \frac{(\Delta + \frac{1}{2}) \Psi(w)}{(z-w)^2} + \frac{\partial \Psi(w)}{z-w} + ... , \\
    &G(z) \Phi(w) = \frac{\Psi(w)}{z-w} + ... , \\
    &G(z) \Psi(w) = \frac{2 \Delta \Phi(w)}{(z-w)^2} + \frac{\partial \Phi(w)}{z-w} + ... ,
\end{align}
where $\Psi$ = $G_{-\frac{1}{2}} \Phi$ -- new field.
Conserved currents can be rewritten in terms of modes $G_r$ and $L_m$. The supercurrent has two forms due to periodic/antiperiodic boundary conditions
\begin{align}
    G^{NS}(z) O(w) = \sum_{r \in \mathbb{Z} + 1/2}  \frac{G_r O(w) }{(z-w)^{r+3/2}},\hspace{20pt} G^{R} (z) O(w) = \sum_{r \in \mathbb{Z}}  \frac{G_r O(w)}{(z-w)^{r+3/2}},
\end{align}
and standard form for the stress-energy tensor
\begin{align}
    T(z) O(w) = \sum_{m \in \mathbb{Z}}  \frac{L_m O(w) }{(z-w)^{m+2}} ,
\end{align}
where $G_r O(w)$, $L_m O(w)$ are just the notation for the new field. Such generators form an algebra \cite{neveu1971factorizable,ramond1971dual} known as the Neveu--Schwarz--Ramond algebra (NSR algebra)
\begin{align} \label{9}
    &[L_m,L_n] = (m-n)L_{m+n} + \frac{\hat{c}}{8} (m^3-m)\delta_{m+n,0}, \nonumber \\
    &[L_m,G_r] = \left( \frac{m}{2} - r \right) G_{m+r},\\
    &\{ G_r,G_s \} = 2L_{r+s} + \frac{\hat{c}}{2}\left( r^2 - \frac{1}{4} \right)\delta_{r+s,0}, \nonumber
\end{align}
where
\begin{align*}
    r,s \in \ &\mathbb{Z} + \frac{1}{2} \text{ for NS-sector}, \\
    r,s \in \ &\mathbb{Z} \hspace{23pt} \text{ for R-sector}. 
\end{align*}

Representation theory of the NSR algebra (\ref{9}) is very similar to the one of the Virasoro algebra. The highest vector corresponding to the primary field must satisfy the following conditions
\begin{align}
    &L_n | \Delta \rangle = G_r | \Delta \rangle =0, \hspace{20pt} n,s > 0 , \\
    &L_0 | \Delta \rangle = \Delta | \Delta \rangle .
\end{align}
In the Ramond sector there are extra conditions due to the zero mode of $G(z)$
\begin{align}
    [G_0, L_0] = 0, \hspace{20pt} G_0^2 = L_0 - \frac{1}{24} \hat{c} .
\end{align}
The Ramond state with the lowest energy is doubly degenerate because both states correspond to the same eigenvalue of $L_0$. In each case eigenvalues $\Delta$ are given as 
\begin{align}
    \Delta_{NS} (\alpha) = \frac{\alpha(Q - \alpha)}{2}, \hspace{20pt} \Delta_R (\alpha) = \Delta_{NS} (\alpha) + \frac{1}{16} .
\end{align}

The set consisting of a primary field and its descendants can be represented using the Verma module $\mathcal{V}_{\Delta}$. Then the conformal family is the linear span of the vectors
\begin{align}
    L_{-\boldsymbol{\lambda}} G_{-\boldsymbol{r}} | \Delta \rangle = L_{-\lambda_1} L_{-\lambda_2} ... G_{-r_1} G_{-r_2} ... | \Delta \rangle,
\end{align}
where $ | \Delta \rangle $ -- highest vector, $\sum \lambda_i + \sum r_j = N$ -- level in the Verma module where the vector is located. There are also conditions for ordering the set $\boldsymbol{\lambda} = \{ \lambda_1 \geq \lambda_2 \geq ...\} $ and strictly ordering the set $\boldsymbol{r} = \{ r_1 \geq r_2 \geq ...\} $. 

A singular vector is a state $| \chi \rangle $ in $\mathcal{V}_{\Delta}$ which is cancelled by the positive part of the Neveu--Schwarz--Ramond algebra
\begin{align}
    L_n | \chi \rangle = G_r | \chi \rangle =0, \hspace{20pt} n,s > 0 .
\end{align}
Without loss of generality, one can assume that the singular vector is an eigenvector of $L_0$
\begin{align}
    L_0 | \chi \rangle = (\Delta + n) | \chi \rangle ,
\end{align}
where n -- level of the singular vector.

For special values of the highest weight and the generic central charge the Verma module is a reducible representation, and it has singular vector. These special values are given by the analog of the Kac-Feigin-Fuchs theorem. The supersymmetric version of this theorem states \cite{poghossian1997structure}, that for
\begin{align}
    \alpha_{m,n} = -\frac{(m-1)b}{2}-\frac{(n-1)b^{-1}}{2}, \hspace{20pt} m,n \in \mathbb{Z}_+ .
\end{align}
There is a singular vector at level $N= \frac{mn}{2}$ which appears in
\begin{align*}
    &m-n \in 2 \mathbb{Z} \hspace{30pt} \text{in R-sector} , \\
    &m-n \in 2 \mathbb{Z}+1 \hspace{10pt} \text{in NS-sector} .
\end{align*}

A singular vector at level $N = \frac{mn}{2}$ corresponds to degenerate field $\Phi_{\alpha}$. Let us consider first examples:
\begin{itemize}
    \item $N=\frac{1}{2}$ - NS-sector 
    \begin{align}
        G_{-\frac{1}{2}} | \Delta \rangle,
    \end{align}
         provided $\Delta = 0$.
    \item $N=1$ - R-sector, two vectors
    \begin{align}
        &\bigg( L_{-1} - \frac{2b^2}{1+2b^2} G_{-1} G_0 \bigg) | \Delta \rangle, \hspace{20pt} \alpha = \alpha_{2,1} = -\frac{b}{2} ,\\
        &\bigg( L_{-1} - \frac{2b^{-2}}{1+2b^{-2}} G_{-1} G_0 \bigg) | \Delta \rangle, \hspace{14pt} \alpha = \alpha_{1,2} = -\frac{1}{2b} .
    \end{align}
    \item $N=\frac{3}{2}$ - NS-sector, two vectors
    \begin{align} \label{221}
        &\bigg(L_{-1} G_{-\frac{1}{2}} + b^2 G_{-\frac{3}{2}} \bigg) | \Delta \rangle, \hspace{20pt} \alpha = \alpha_{3,1} = -b , \\
        &\bigg(L_{-1} G_{-\frac{1}{2}} + b^{-2} G_{-\frac{3}{2}} \bigg) | \Delta \rangle, \hspace{14pt} \alpha = \alpha_{3,1} = -b^{-1} . \label{22}
    \end{align}
\end{itemize}

\section{Ward identities in superconformal field theories} \label{sec: Chapter 3}
Conformal symmetry imposes certain constraints on correlation functions called global Ward Identities \cite{belavin1984infinite}. Moreover, each conserved current has its own Ward identity. These conditions impose special restrictions that are related to the fact that correlation functions tend to zero at infinity. They play an important role, since they provide explicit connections between correlation functions with different sets of fields.

The equation corresponding to the energy-momentum tensor has a fairly well-known form
\begin{align} \label{23}
    \langle T(z) \Phi_1(z_1) ... \Phi_n(z_n) \rangle =& \sum_{k=1}^n \bigg( \frac{\Delta_k}{(z-z_k)^2} + \frac{\partial_k}{z-z_k} \bigg) \langle \Phi_1(z_1) ... \Phi_n(z_n) \rangle , 
\end{align}
and similar expressions including the field $\Psi(z)$. It is known that the correlation function including $T(z)$ must fall off at infinity as
\begin{align}
    \langle T(z) ... \rangle \thicksim \frac{1}{z^4}, \hspace{20pt} \text{at } z \rightarrow \infty .
\end{align}
Therefore, the terms of order $1/z,1/z^2$ and $1/z^3$ on the right-hand side of (\ref{23}) must vanish
\begin{align} \label{(3.4)}
    &\sum_{k=1}^n \partial_k \langle \Phi_1(z_1) ... \Phi_n(z_n) \rangle = 0 , \nonumber \\
    &\sum_{k=1}^n (\Delta_k + z_k \partial_k) \langle \Phi_1(z_1) ... \Phi_n(z_n) \rangle = 0 , \\
    &\sum_{k=1}^n (2 z_ k \Delta_k + z_k^2 \partial_k) \langle \Phi_1(z_1) ... \Phi_n(z_n) \rangle = 0 . \nonumber
\end{align}

To understand the structure of global Ward identities for the supercurrent $G(z)$, we consider the example of two-point correlation function. Thus, for the NS-sector, the equations take the form
\begin{align*}
    &\langle G(z) \Phi_1(z_1) \Phi_2(z_2) \rangle = \frac{\langle \Psi_1 (z_1) \Phi_2 (z_2) \rangle}{z-z_1} + \frac{\langle \Phi_1 (z_1) \Psi_2 (z_2) \rangle}{z-z_2} , \\
    &\langle G(z) \Phi_1(z_1) \Psi_2(z_2) \rangle = \frac{\langle \Psi_1 (z_1) \Psi_2 (z_2) \rangle}{z-z_1} + \frac{2 \Delta_2 \langle \Phi_1(z_1) \Phi_2(z_2) \rangle}{(z-z_2)^2} + \frac{\partial_{z_2} \langle \Phi_1(z_1) \Phi_2(z_2) \rangle}{z-z_2} .
\end{align*}
In contrast to the equation (\ref{23}) for $T(z)$, the insertion of  $G(z)$ produces correlation functions of different sets of fields. This fact is expected because there are two fields, $\Phi(z)$ and $\Psi(z)$, which interact differently with  $G(z)$. The generalization for $n$-point correlation functions
\begin{align} \label{24}
    \langle G(z) ... \Phi_i(z_i) ... \Psi_j(w_j) ... &\rangle = \sum_{i=1}^n (-1)^{k_i} \frac{\langle ... \Psi_i(z_i) ... \Psi_j(w_j) ... \rangle}{z-z_i} + \nonumber \\
    &+ \sum_{j=1}^m (-1)^{j-1} \bigg( \frac{2 \Delta_j}{(z-w_j)^2} + \frac{\partial_{w_j} }{z-w_j} \bigg) \langle ... \Phi_i(z_i) ... \Phi_j(w_j) ... \rangle,
\end{align}
where i belongs to bosons, and j belongs to fermions. There is $k_i$ in the first line that shows the number of fermions before $\Phi_i(z_i)$.

Similarly to  $T(z)$, the supercurrent $G(z)$ should falls at infinity as
\begin{align}
    \langle G(z) ... \rangle \thicksim \frac{1}{z^3}, \hspace{20pt} \text{at } z \rightarrow \infty .
\end{align}
General expressions similar to (\ref{(3.4)}) are huge for supercurrent case, and so we give below a special case of four-point functions with two fermions on the right side. The disappearance of orders $1/z,1/z^2$ for (\ref{24}) is ensured through the following equations
\begin{align} \label{(3.6)}
\begin{split}
         &\underset{i \neq j}{\underset{i=1}{\sum^4}} (-1)^{\theta(j-i)} \langle ... \Psi_i(z_i) ... \Psi_j(w_j) ... \rangle + \partial_j \langle ... \Phi_j(w_j)  ... \rangle = 0 , \\
    &\underset{i \neq j}{\underset{i=1}{\sum^4}} (-1)^{\theta(j-i)} z_i \langle ... \Psi_i(z_i) ... \Psi_j(w_j) ... \rangle + (2 \Delta_j + z_j \partial_j) \langle ... \Phi_j(w_j)  ... \rangle =0,
\end{split}
\end{align}
where $\theta(j-i)$ -- Heaviside step function.

Consequences from the Ward Identities for the stress-energy tensor are used to express the derivatives. In calculations, it is convenient to represent all derivatives through one of them. Thus we present explicit expressions for the derivatives in terms of $\partial$, related to the first field in $\langle O_{\alpha}(z) O_1(z_1) O_2(z_2) O_3(z_3) \rangle$
\begin{align}
    \partial_i = \frac{- (z-z_k)(z-z_l) \partial + \underset{m=0,i}{\sum} \Delta_m (-2z_m + \underset{j=k,l}{\sum} z_j) +  (- \Delta_k + \Delta_l) (z_k - z_l)}{(z_i - z_k)(z_i - z_l)}, 
\end{align}
where $k,l \neq i$;  $k<l$; $k,l \in \overline{1,3}$. 
The constraints connected with the supercurrent are used to relate the correlation functions of different numbers of fermions. In particular, these relations allow us to express functions through each other. To see how it works, let us consider the subsystem containing correlation functions with an even number of fermions. The goal is to express these functions through the bosonic one. The restrictions on these correlation functions are obtained from the Ward identities on $\langle G \Psi_{\alpha} \Phi_1 \Phi_2 \Phi_3 \rangle$, $\langle G \Phi_{\alpha} \Psi_1 \Phi_2 \Phi_3 \rangle$, $\langle G \Phi_{\alpha} \Phi_1 \Psi_2 \Phi_3 \rangle$, $\langle G \Phi_{\alpha} \Phi_1 \Phi_2 \Psi_3 \rangle$. It is easy to note that a closed system of eight equations can be chosen for the following seven functions\footnote{In the end of the procedure, the fermion function $\langle \Psi_{\alpha}(z) \Psi_1(z_1) \Psi_2(z_2) \Psi_3(z_3) \rangle $ can be easily expressed through bosonic one, since it is related to the correlation functions with two fermions via the Ward identities.}
\begin{align} \label{300}
\begin{split}
       \langle \Phi_{\alpha} \Phi_1 \Phi_2 \Phi_3 \rangle, \langle \Psi_{\alpha} \Psi_1 \Phi_2 \Phi_3 \rangle,& \langle \Psi_{\alpha} \Phi_1 \Psi_2 \Phi_3 \rangle, \langle \Psi_{\alpha} \Phi_1 \Phi_2 \Psi_3 \rangle, \\
        \langle \Phi_{\alpha} \Psi_1 \Psi_2 \Phi_3 \rangle,\langle \Phi_{\alpha} \Psi_1& \Phi_2 \Psi_3 \rangle, \langle \Phi_{\alpha} \Phi_1 \Psi_2 \Psi_3 \rangle .
\end{split}
\end{align}
Further, using restrictions related to Ward identities, it is easy to express all these functions through $\langle \Phi_{\alpha} \Phi_1 \Phi_2 \Phi_3 \rangle $ and $\langle \Phi_{\alpha} \Psi_1 \Psi_2 \Phi_3 \rangle$. The reason why there are not enough equations in the system to express all correlation functions is due to the fact that it contains hidden Ward identities (\ref{(3.4)}) for the bosonic function.

To obtain an extra equation, an additional condition must be satisfied. Let us assume that the first field is degenerate in the correlation function, and $\alpha = -b$. In this case, according to (\ref{221}) we have a singular vector at level $N=3/2$
\begin{align} \label{(3.8)}
    (L_{-1} G_{-\frac{1}{2}} + b^2 G_{-\frac{3}{2}}) | \Delta \rangle .
\end{align}
This vector and its descendants allow not only to express correlation functions through each other, but also to derive differential equations on them. In particular, additional conditions are obtained with the help of the following relations
\begin{align}
    \langle L_{-k} \Phi(z) \Phi_1(z_1) ... \Phi_n(z_n) \rangle =& \sum_{j=1}^n \bigg( \frac{\Delta_j (k-1)}{(z_j-z)^k} - \frac{\partial_j}{(z_j-z)^{k-1}} \bigg) \langle \Phi(z) \Phi_1(z_1) ... \Phi_n(z_n) \rangle , \\
    \langle G_{-r} O(z)... \Phi_i(z_i) ... \Psi_j(w_j) ... \rangle =& -\sum_{i=1}^n (-1)^{k_i} \frac{\langle ... \Psi_i(z_i) ... \Psi_j(w_j) ... \rangle}{(z_i-z)^{r-1/2}} + \nonumber \\
    + \sum_{j=1}^m (-1)^{j-1} \bigg( &\frac{2 \Delta_j   (r-\frac{1}{2})}{(w_j-z)^{r+1/2}} - \frac{\partial_{w_j} }{(w_j-z)^{r-1/2}} \bigg) \langle ... \Phi_i(z_i) ... \Phi_j(w_j) ... \rangle ,
\end{align}
where $O(z)$ -- boson/fermion field, added to the numeration in the corresponding sum. 

In particular, the new equation from the descendant of the singular vector 
\begin{align}
    G_{-\frac{1}{2}} (L_{-1} G_{-\frac{1}{2}} + b^2 G_{-\frac{3}{2}}) | \Delta \rangle 
\end{align}
allows to express all correlation functions in (\ref{300}) via the bosonic function $\langle \Phi_{-b}(z) \Phi_1(z_1) \Phi_2(z_2) \Phi_3(z_3) \rangle$. Thus, expressions for all functions are found, including $\langle \Psi_{-b}(z) \Psi_1(z_1) \Psi_2(z_2) \Psi_3(z_3) \rangle$.

This procedure of finding correlation functions is similar for the odd subsystem
\begin{align} \label{301}
\begin{split}
       &\langle \Psi_{\alpha} \Phi_1 \Phi_2 \Phi_3 \rangle, \langle \Phi_{\alpha} \Psi_1 \Phi_2 \Phi_3 \rangle, \langle \Phi_{\alpha} \Phi_1 \Psi_2 \Phi_3 \rangle, \langle \Phi_{\alpha} \Phi_1 \Phi_2 \Psi_3 \rangle, \\
       &\langle \Phi_{\alpha} \Psi_1 \Psi_2 \Psi_3 \rangle, \langle \Psi_{\alpha} \Phi_1 \Psi_2 \Psi_3 \rangle, \langle \Psi_{\alpha} \Psi_1 \Phi_2 \Psi_3 \rangle, \langle \Psi_{\alpha} \Psi_1 \Psi_2 \Phi_3 \rangle .
\end{split}
\end{align}
In this case we construct additional equation using 
\begin{align}
    L_{-1}(L_{-1} G_{-\frac{1}{2}} + b^2 G_{-\frac{3}{2}}) | \Delta \rangle 
\end{align}  
and express functions via $\langle \Psi_{-b}(z) \Phi_1(z_1) \Phi_2(z_2) \Phi_3(z_3) \rangle$. We present explicit expressions for all correlation functions in \hyperref[sec: Apendix A]{Apendix A}. 

\section{Differential equation for the correlation function} \label{sec: Chapter 4}
As it is already known, all four-point correlation functions can be divided into two subsystems (\ref{300}) and (\ref{301}). In each of them, we can select the basic function and find it as a solution of the special differential equation. Firstly, considering the even subsystem, we will find the value for $\langle \Phi_{-b}(z) \Phi_1(z_1) \Phi_2(z_2) \Phi_3(z_3) \rangle$. The procedure is similar for the odd subsystem, but basic function will be $\langle \Psi_{-b}(z) \Phi_1(z_1) \Phi_2(z_2) \Phi_3(z_3) \rangle$. 

To construct a differential equation for bosonic function, we turn to the next descendant of singular vector (\ref{(3.8)})
\begin{align} \label{435}
    G_{-\frac{3}{2}}(L_{-1} G_{-\frac{1}{2}} + b^2 G_{-\frac{3}{2}}) | \Delta \rangle .
\end{align}
Further, we will consider only the holomorphic part, and discussion for the antiholomorphic part is similar. Using the descendant (\ref{435}), we obtain a differential equation on bosonic correlation function
\begin{align*}
    b^2 \sum_{i=1}^3 \bigg( \frac{2 \Delta_i}{(z-z_i)^3} + \frac{\partial_i}{(z-z_i)^2} \bigg) \langle \Phi(z) \Phi_1(z_1) \Phi_2(z_2) \Phi_3(z_3) \rangle + \sum_{i=1}^3 \bigg( \frac{1}{z-z_i} \partial_z \langle \Psi(z) ... \Psi_i(z_i) ... \rangle \bigg) = 0 .
\end{align*}

We can eliminate functions with fermions by using the expressions from \hyperref[sec: Apendix A]{Apendix A}. Projective invariance allows to fix $(z, z_1, z_2, z_3) \rightarrow (w,0,1, \infty) $. After a special substitution $y (w) = \langle \Phi_{-b}(w) \Phi_1(0) \Phi_2(1) \Phi_3(\infty) \rangle $, we obtain a simplified form of the supersymmetric version of BPZ equation:
\begin{align}
    &\frac{1}{b^2} y'''(w) + \frac{(-1 + 2 b^2) (1 - 2 w)}{b^2 (w-1) w} y''(w) + \nonumber \\
    &\bigg( \frac{b^2 (1 + 3 (-1 + w) w) + 2 \Delta_1 + w (3 + 2 \Delta - 2 \Delta_1 + 
    2 \Delta_2 - 2 \Delta_3 + 
    w (-3 - 2 \Delta + 2 \Delta_3)) }{(w-1)^2 w^2} \bigg) y'(w) + \nonumber \\
    &\bigg( \frac{2 ((-1 + w)^2 \Delta_1 - w^2 \Delta_2) + 
 b^2 (2 \Delta_1 + 
    2 w^3 (\Delta - \Delta_3) + 
    w (\Delta - 
       5 \Delta_1 + \Delta_2 - \Delta_3)}{(w-1)^3 w^3} - \\  
       &- \frac{ 3 w^2 (\Delta - \Delta_1 + \Delta_2 - \Delta_3))}{(w-1)^3 w^3} \bigg) y(w) = 0 . \nonumber
\end{align}
This equation can be reduced to hypergeometric type form using the following substitution
\begin{align} \label{36}
    y(w) = w^{\alpha_1 b} (1-w)^{\alpha_2 b} F(w) .
\end{align}
After that, we obtain the Dotsenko-Fateev equation \cite{dotsenko1984conformal}
\begin{align} \label{37}
\begin{split}
    &w^2 (w-1)^2 F'''(w) + \big( K_1 w + K_2 (w-1) \big) w (w-1) F''(w) + \\
    &+\big( L_1 w^2 + L_2 (w - 1)^2 + L_3 w(w-1) \big) F'(w) + (M_1 w + M_2 (w-1)) F(w) = 0,
\end{split}
\end{align}
where the parameters $K_i$, $L_i$ and $M_i$ are given in \hyperref[sec: Apendix B]{Apendix B}. It is important to note that this is not true hypergeometric equation, but it can be reduced to one. In case $L_1 = M_1 = 0$ we have equation that correspond to $_3 F_2 (a_1, a_2, a_3;b_1, b_2;w)$.

Below we present the results that follow from (\ref{37}). A brief but more detailed description from \cite{dotsenko1984conformal} is given in \hyperref[sec: Apendix B]{Apendix B}. Solutions of (\ref{37}) have the form of Coulomb gas integral \cite{dotsenko1984conformal}
\begin{align} \label{38}
    I(A,B,C,g;w) = \int\limits_{C1} dt_1 \int\limits_{C2} dt_2 t_1^A (t_1 - 1)^B (t_1 - w)^C  t_2^A (t_2 - 1)^B (t_2 - w)^C (t_1 - t_2)^g .
\end{align}
where the set of parameters $\{ A,B,C,g \}$ can take any of two sets 
\begin{align*}
    &A_1 = \frac{1}{2} b (b - \alpha_1 + \alpha_2 - \alpha_3), \hspace{20pt} B_1 =\frac{1}{2} b (b + \alpha_1 - \alpha_2 - \alpha_3), \\
    &C_1 = \frac{1}{2} b (b - \alpha_1 - \alpha_2 + \alpha_3), \hspace{21pt} g_1 = -1 - b^2;  \\ 
    &A_2 = \frac{1}{2} \big(-1 + b (-\alpha_1 + \alpha_2 + \alpha_3) \big), \hspace{20pt} 
    B_2 = \frac{1}{2} \big(-1 + b (\alpha_1 - \alpha_2 + \alpha_3) \big), \\
    &C_2 = \frac{1}{2} + b^2 -\frac{1}{2} b (\alpha_1 + \alpha_2 + \alpha_3), \hspace{30pt} g_2 = -1 - b^2 .
\end{align*}
This is due to the fact that there is a reflection relation in SLFT for the fields 
\begin{align} \label{39}
    &\Phi_{\alpha} = R(\alpha) \Phi_{Q - \alpha}, 
\end{align}
where
\begin{align} \label{40}
    R(\alpha) = \frac{(\pi \mu \gamma(b^2))^{\frac{2\alpha - Q}{b}}}{b^2} \frac{\gamma( 2b \alpha -b^2)}{\gamma(2 - 2b^{-1} \alpha + b^{-2})} .
\end{align}
Thus, one of sets $\{ A,B,C,g \}$ corresponds to $\Phi_{\alpha}$, and the other corresponds to $\Phi_{Q-\alpha}$. These two sets are transformed into each other by replacing $\alpha_3 \rightarrow Q - \alpha_3$. Therefore, any set can be chosen, and the difference for related correlation functions will be expressed in a numerical factor. In this work, we choose and fix the first set.

To form the physical correlation function, we have to restore the dependence on $\bar{w}$. The physical correlation function should be of the following form \cite{dotsenko1984conformal} : 
\begin{align} \label{41}
    G(w, \Bar{w}) = X_{1} |I_1(w)|^2 + X_{2} |I_2(w)|^2 + X_{3} |I_3(w)|^2, 
\end{align}
where
\begin{align}
    I_1(w) = \int\limits_{1}^{\infty} dt_1 \int\limits_{1}^{\infty} dt_2 t_1^A (t_1 - 1)^B (t_1 - w)^C  t_2^A (t_2 - 1)^B (t_2 - w)^C (t_1 - t_2)^g , \\
    I_2(w) = \int\limits_{1}^{\infty} dt_1 \int\limits_{0}^{w} dt_2 t_1^A (t_1 - 1)^B (t_1 - w)^C  t_2^A (t_2 - 1)^B (t_2 - w)^C (t_1 - t_2)^g , \\
    I_3(w) = \int\limits_{0}^{w} dt_1 \int\limits_{0}^{w} dt_2 t_1^A (t_1 - 1)^B (t_1 - w)^C  t_2^A (t_2 - 1)^B (t_2 - w)^C (t_1 - t_2)^g .
\end{align}
Also, the ratios of coefficient $X_i$ are fixed from the condition that the correlation function is single-valuated 
\begin{align} 
    \frac{X_1}{X_3} = &- \frac{\sin{\pi (1 + b \alpha_1)}  \sin{ \frac{\pi}{2} (-1 + b (\alpha_1 - \alpha_2 - \alpha_3))} \sin{ \frac{\pi b}{2} (b + \alpha_1 - \alpha_2 - \alpha_3) } }{\sin{b \pi (b - \alpha_1)}  \sin{\frac{\pi}{2} (1 + b (\alpha_1 + \alpha_2 - \alpha_3)) } \sin{ \frac{\pi b}{2} (b - \alpha_1 + \alpha_2 - \alpha_3) } } \cdot \nonumber\\
    & \cdot \frac{\sin{\frac{\pi}{2} (-2 + b^2 - b (\alpha_1 + \alpha_2 + \alpha_3)) } \sin{ \frac{\pi}{2} (-1 + 2 b^2 - b (\alpha_1 + \alpha_2 + \alpha_3))}}{\sin{ \frac{\pi b}{2} (b - \alpha_1 - \alpha_2 + \alpha_3) } \sin{\frac{\pi}{2} (1 + b (\alpha_1 - \alpha_2 + \alpha_3))}} , \\
    \frac{X_2}{X_3} = & \frac{ \sin{ \frac{\pi}{2} (1 + b^2) } \sin{\frac{\pi}{2} (-1 + b^2 - 2 b \alpha_1) } \sin{ \frac{\pi b}{2} (b + \alpha_1 - \alpha_2 - \alpha_3) }}{\sin{ \pi (1 + b^2) } \sin{
  b \pi (b - \alpha_1) } \sin{\frac{\pi}{2} (1 + b (\alpha_1 + \alpha_2 - \alpha_3)) }} \cdot \nonumber\\
    & \cdot \frac{\sin{ \frac{\pi}{2} (-2 + b^2 - b (\alpha_1 + \alpha_2 + \alpha_3))}}{\sin{ \frac{\pi}{2} (1 + b (\alpha_1 - \alpha_2 + \alpha_3))}} .
\end{align}

Similarly, we can consider the correlation function $\langle \Psi_{-b} \Phi_1 \Phi_2 \Phi_3 \rangle$. The vector $G_{-\frac{1}{2}}G_{-\frac{3}{2}}(L_{-1} G_{-\frac{1}{2}} + b^2 G_{-\frac{3}{2}}) | \Delta \rangle$ allows to construct another differential equation
\begin{align}
    &\frac{1}{b^2} y'''(w) - \frac{2 (1 - b^2) (1 - 2 w)}{b^2 (w - 1) w} y''(w) + \bigg( \frac{2}{b^2 (w - 1) w} + \nonumber \\
    &\frac{ b^4 (1 - 3 w + 3 w^2) + 
    b^2 (-1 + 2 \Delta_1 - 
    2 w^2 (4 + \Delta - \Delta_3) + 
    2 w (4 + \Delta - \Delta_1 + \Delta_2 - \Delta_3)))}{b^2 (w - 1)^2 w^2} \bigg) y'(w) + \nonumber \\
    &+ b^2 \bigg( \frac{4 \Delta_1 + 
    w^3 (6 + 4 \Delta - 4 \Delta_3) + 
    w (3 + 2 \Delta - 10 \Delta_1 + 
    2 \Delta_2 - 2 \Delta_3)}{2 (w - 1)^3 w^3} +\\
    &+ \frac{w^2 (-9 - 6 \Delta + 6 \Delta_1 - 
    6 \Delta_2 + 6 \Delta_3)}{2 (w - 1)^3 w^3} \bigg) y(w) = 0 . \nonumber
\end{align}
Its solution is also integral (\ref{38}) with two sets of parameters 
\begin{align*}
    &A_1 = -1 + \frac{1}{2} b (-\alpha_1 + \alpha_2 + \alpha_3), \hspace{20pt} 
    B_1 = -1 + \frac{1}{2} b (\alpha_1 - \alpha_2 + \alpha_3), \\
    &C_1 = \frac{1}{2} b (2 b - \alpha_1 - \alpha_2 - \alpha_3), \hspace{35pt}  
    g = 1 - b^2; \\
    &A_2 =\frac{1}{2} (-1 + b (b - \alpha_1 + \alpha_2 - \alpha_3)), \hspace{20pt} B_2 = \frac{1}{2} (-1 + b (b + \alpha_1 - \alpha_2 - \alpha_3)), \\
    &C_2 = \frac{1}{2} (-1 + b (b - \alpha_1 - \alpha_2 + \alpha_3)), \hspace{12pt} g = 1 - b^2 .
\end{align*}
After fixing the first set, we find 
\begin{align}
    \frac{X_1}{X_3} =& \frac{\sin{b \pi \alpha_1 } \sin{ \frac{\pi}{2} (-3 + b^2 - b (\alpha_1 + \alpha_2 - \alpha_3)) } \sin{\frac{\pi }{2} (2 + b (\alpha_1 + \alpha_2 - \alpha_3)) }}{\sin{\pi (-1 + b^2 - b \alpha_1) } \sin{\frac{\pi}{2} (1 + b^2 + b (\alpha_1 - \alpha_2 - \alpha_3)) } \sin{ \frac{\pi b}{2} (2 b - \alpha_1 - \alpha_2 - \alpha_3) }} \cdot \nonumber\\
    & \cdot \frac{\sin{ \frac{\pi}{2} (1 + b^2 -
    b (\alpha_1 - \alpha_2 + \alpha_3)) } sin{\pi (-1 + \frac{1}{2} b (\alpha_1 - \alpha_2 + \alpha_3))}}{\sin{\pi (-1 + \frac{1}{2} b (-\alpha_1 + \alpha_2 + \alpha_3)) } \sin{ \frac{\pi}{2} (1 + b^2 - b (\alpha_1 + \alpha_2 + \alpha_3))}} , \\
    \frac{X_2}{X_3} =& \frac{ \sin{\frac{\pi}{2} (-1 + b^2) }  \sin{ \frac{\pi}{2} (-1 + b^2 - 2 b \alpha_1) } \sin{\frac{\pi}{2} (2 + b (\alpha_1 + \alpha_2 - \alpha_3)) }}{ \sin{\pi (-1 + b^2 ) } \sin{\pi (-1 + b^2 - b \alpha_1) } \sin{ \frac{\pi}{2} (1 + b^2 + b (\alpha_1 - \alpha_2 - \alpha_3)) }} \cdot \nonumber\\
    & \cdot \frac{\sin{\pi (-1 + 
     \frac{1}{2} b (\alpha_1 - \alpha_2 + \alpha_3))}}{\sin{
  \frac{\pi}{2} (1 + b^2 - b (\alpha_1 + \alpha_2 + \alpha_3))}} .
\end{align}

\section{Superconformal operator product expansions}
In the previous section, it was shown that the correlation functions with degenerate field $\Phi_{-b}$ admit an integral representation. These results should be compared to general restrictions on operator product expansion (OPE) in CFT. In particular, one can use the so called Teschner's trick, mentioned in introduction, to find the structure constants of the OPE.

It is known that the field space in CFT is decomposed into a direct sum of representations of the Virasoro algebra
\begin{align}
    V = \underset{\alpha}{\oplus} [ O_{\alpha} ],
\end{align}
where $[ O_{\alpha} ]$ - conformal field. Each family consists of a primary field $O_{\alpha}$ and all its descendants obtained by repeatedly applying to $O_{\alpha}$ the generators $L_n$ and $G_r$, $n,s < 0$.

It turns out that there is a restriction on the OPE of the degenerate field with others. Let us consider the expansion using the example of two boson fields located at the points $z$ and $0$. For $z \rightarrow 0$, we obtain the following expansion
\begin{align}
    \Phi_{\alpha} (w, \Bar{w}) \Phi_{\alpha_1} (0) = \sum_s C_{\alpha, \alpha_1}^{s} |w|^{2(\Delta_s - \Delta_{-b} - \Delta_{\alpha_1})} O_{s} (0) + ... ,
\end{align}
where the field $O_s$ can be either bosonic or fermionic\footnote{ Previously we considered the holomorphic part of the function and adopted the notation $\Psi = G_{- \frac{1}{2}} \Phi $, but now $\Psi = G_{- \frac{1}{2}} \bar{G}_{- \frac{1}{2}} \Phi $ }. Applying this general information to the special case considered in \hyperref[sec: Chapter 4]{section 4}, we find an expansion for the fields $\Phi_{-b} (w)$ and $\Phi_{\alpha_1} (0)$. As a result, three classes arise: two bosonic and one fermionic. Let us represent this schematically
\begin{align} \label{53}
    [\Phi_{-b}]  [\Phi_{\alpha_1}] = [\Phi_{\alpha_1-b}] + [\Psi_{\alpha_1}] + [\Phi_{\alpha_1+b}] .
\end{align}

Let us recall that three-point correlation functions depend on coordinates $(z_k , \bar{z}_k)$ in an universal way
\begin{align}
    \langle \Phi_{\alpha_1} (z_1, \Bar{z}_1) \Phi_{\alpha_2} (z_2, \Bar{z}_2) \Phi_{\alpha_3} (z_3, \Bar{z}_3)  \rangle =& C(\alpha_1,\alpha_2,\alpha_3) \prod_{i < j} |z_i - z_j|^{-2 \Delta_{ij}} ,\\
    \langle \Psi_{\alpha_1} (z_1, \Bar{z}_1) \Phi_{\alpha_2} (z_2, \Bar{z}_2) \Phi_{\alpha_3} (z_3, \Bar{z}_3)  \rangle =& \widetilde{C}(\alpha_1,\alpha_2,\alpha_3) \prod_{i < j} |z_i - z_j|^{-2 \Delta_{ij}} ,
\end{align}
where $\Delta_{ij} = \Delta_i + \Delta_j - \Delta_k $. The choice of points $(z_1, z_2, z_3) = (0,1, \infty) $ fixes the structure constants
\begin{align}
    &\langle \Phi_{\alpha_1} (0) \Phi_{\alpha_2} (1) \Phi_{\alpha_3} (\infty)  \rangle = C(\alpha_1,\alpha_2,\alpha_3) , \\
    &\langle \Psi_{\alpha_1} (0) \Phi_{\alpha_2} (1) \Phi_{\alpha_3} (\infty)  \rangle = \widetilde{C} (\alpha_1,\alpha_2,\alpha_3) .
\end{align}

General expansion for the four-point boson correlation function can be rewritten using the structure constants mentioned above
\begin{align} \label{58}
    \langle \Phi_{-b} (w, \Bar{w}) \Phi_{\alpha_1} (0) \Phi_{\alpha_2} (1)& \Phi_{\alpha_3} (\infty) \rangle = \nonumber \\
    &\sum_s C_{-b, \alpha_1}^{s} |w|^{2(\Delta_s - \Delta_{-b} - \Delta_{\alpha_1})} \langle \Phi_{s} (0) \Phi_{\alpha_2} (1) \Phi_{\alpha_3} (\infty) \rangle + ... \nonumber \\ 
    &+ \sum_s \widetilde{C}_{-b, \alpha_1}^{s} |w|^{2(\Delta_s + \frac{1}{2} - \Delta_{-b} - \Delta_{\alpha_1})} \langle \Psi_{s} (0) \Phi_{\alpha_2} (1) \Phi_{\alpha_3} (\infty) \rangle + ...\nonumber \\ 
    =& \sum_s C_{-b, \alpha_1}^{s} C(s,\alpha_2,\alpha_3) |w|^{2(\Delta_s - \Delta_{-b} - \Delta_{\alpha_1})} + ...  \nonumber \\
    &+ \sum_s \widetilde{C}_{-b, \alpha_1}^{s} \widetilde{C}(s,\alpha_2,\alpha_3) |w|^{2(\Delta_s + \frac{1}{2} - \Delta_{-b} - \Delta_{\alpha_1})} + ... ,
\end{align}
where $s = \alpha - b, \alpha + b$ in the first sum;  $s = \alpha$ in the second sum.

We can consider this situation from the other side and turn to the obtained solutions (\ref{41}). We expand the integrals $I_k(w)$ near $w=0$. For $w \rightarrow 0$ we obtain
\begin{align} \label{59}
    &I_1(w) = {\cal N}_1 (1 + a_1 w + ...) , \nonumber \\
    &I_2(w) = {\cal N}_2 w^{(1 + A + C)} (1 + a_2 w + ...) , \\
    &I_3(w) = {\cal N}_3 w^{2(1 + A + C) + g} (1 + a_3 w + ...) , \nonumber
\end{align}
where the normalization constants ${\cal N}_i$ 
\begin{align} \label{(6.10)}
    &{\cal N}_1 = \frac{\Gamma(g) \Gamma(-1-A-B-C-g) \Gamma(-1-A-B-C-\frac{1}{2}g) \Gamma(1+B) \Gamma(1+B+ \frac{1}{2}g)}{\Gamma(\frac{1}{2}g) \Gamma(-A-C) \Gamma(-A-C-\frac{1}{2}g)} , \nonumber \\
    &{\cal N}_2 = \frac{\Gamma(-1-A-B-C-g) \Gamma(1+B) \Gamma(1+A) \Gamma(1+C)}{\Gamma(-A-C-g) \Gamma(2+A+C)} , \\
    &{\cal N}_3 = \frac{\Gamma(g) \Gamma(1+A) \Gamma(1+A+ \frac{1}{2}g) \Gamma(1+C) \Gamma(1+C+ \frac{1}{2}g)}{\Gamma(\frac{1}{2}g) \Gamma(2+A+C+ \frac{1}{2}g) \Gamma(2+A+C+g)} \nonumber 
\end{align}
can be calculated using the Selberg integral \cite{selberg1944remarks}
\begin{align*}
    \int \prod \limits_{i=1}^n t_i^{\alpha-1} (1-t_i)^{\beta-1} \prod \limits_{i<j} |t_i - t_j|^{2 \gamma} dt_1 ... dt_n =\prod \limits_{j=0}^{n-1} \frac{\Gamma(\alpha +j \gamma) \Gamma(\beta +j \gamma) \Gamma(1 + (j+1) \gamma)}{\Gamma(\alpha + \beta + (n+j-1)\gamma) \Gamma(1+ \gamma)} .
\end{align*}
Comparing asymptotic at $w \rightarrow 0$, it can be shown that the following identification holds
\begin{align*}
    [\Phi_{\alpha_1-b}] \rightarrow I_1(w), \hspace{20pt} [\Psi_{\alpha_1}] \rightarrow I_2(w), \hspace{20pt} [\Phi_{\alpha_1+b}] \rightarrow I_3(w) .
\end{align*}

Considering (\ref{59}), the expression for the correlation function $G(w, \Bar{w})$ (\ref{41}) can be rewritten using the $X_i$ factors and normalization constants
\begin{align}
    G(w, \Bar{w}) = X_{1} {\cal N}_1^2 |I_1(w)|^2 + X_{2} {\cal N}_2^2 |I_2(w)|^2 + X_{3} {\cal N}_3^2 |I_3(w)|^2 .
\end{align}
Comparing with (\ref{58}), we find
\begin{align}
\begin{split}
    C_{-b, \alpha_1}^{\alpha_1 - b} C(\alpha_1 - b,\alpha_2,\alpha_3) = X_{1} {\cal N}_1^2 , &\hspace{20pt} C_{-b, \alpha_1}^{\alpha_1 + b} C(\alpha_1 + b,\alpha_2,\alpha_3) = X_{3} {\cal N}_3^2 , \\
    \widetilde{C}_{-b, \alpha_1}^{\alpha_1} \widetilde{C}(\alpha_1,&\alpha_2,\alpha_3) = X_{2} {\cal N}_2^2 ,
\end{split}
\end{align}
which is equivalent to functional equations 
\begin{align}
    \frac{C_{-b, \alpha_1}^{\alpha_1 - b} C(\alpha_1 - b,\alpha_2,\alpha_3)}{C_{-b, \alpha_1}^{\alpha_1 + b} C(\alpha_1 + b,\alpha_2,\alpha_3)}= \frac{X_{1} {\cal N}_1^2}{X_{3} {\cal N}_3^2} , \\
    \frac{ \widetilde{C}_{-b, \alpha_1}^{\alpha_1} \widetilde{C}(\alpha_1, \alpha_2,\alpha_3) }{C_{-b, \alpha_1}^{\alpha_1 + b} C(\alpha_1 + b,\alpha_2,\alpha_3)}= \frac{X_{2} {\cal N}_2^2}{X_{3} {\cal N}_3^2} ,
\end{align}
or explicitly
\begin{align}
    \frac{C_{-b, \alpha_1}^{\alpha_1 - b} C(\alpha_1 - b,\alpha_2,\alpha_3)}{C_{-b, \alpha_1}^{\alpha_1 + b} C(\alpha_1 + b,\alpha_2,\alpha_3)} = &\frac{1}{4} b^2 \alpha_1^2 (1 + b^2 - 2 b \alpha_1)^2 
    \frac{\gamma\big(\frac{1}{2} (1 + b (\alpha_1 + \alpha_2 - \alpha_3))\big) \gamma\big(\frac{1}{2} b (-b + \alpha_1 + \alpha_2 - \alpha_3)\big) }{\gamma\big(b (-b + \alpha_1)\big) \gamma\big(1 + b \alpha_1\big)} \cdot \nonumber \\ 
    &\cdot \frac{\gamma\big(\frac{1}{2} b (-b + \alpha_1 - \alpha_2 + \alpha_3)\big) \gamma\big( \frac{1}{2} b (-b + \alpha_1 + \alpha_2 + \alpha_3)\big) }{\gamma\big( \frac{1}{2} - b^2/2 + b \alpha_1\big)^2 \gamma\big( \frac{1}{2} b (-b - \alpha_1 + \alpha_2 + \alpha_3)\big)} \cdot \nonumber\\ 
    & \cdot \frac{\gamma\big( \frac{1}{2} (1 + b (\alpha_1 - \alpha_2 + \alpha_3) \gamma\big(\frac{1}{2} (-1 + b (-2 b + \alpha_1 + \alpha_2 + \alpha_3))\big) }{\gamma\big( \frac{1}{2} (1 + b (-\alpha_1 + \alpha_2 + \alpha_3))\big)} , \label{64} \\ 
    \frac{ \widetilde{C}_{-b, \alpha_1}^{\alpha_1} \widetilde{C}(\alpha_1, \alpha_2,\alpha_3) }{C_{-b, \alpha_1}^{\alpha_1 + b} C(\alpha_1 + b,\alpha_2,\alpha_3)} = & -\frac{b^2 \alpha_1^2 (1 + b^2 - 2 b \alpha_1)^2}{(1 + b^2 - b \alpha_1)^2 } \frac{\gamma\big(1 + b^2\big) \gamma\big(b (-b + \alpha_1)\big) \gamma\big(\frac{1}{2} (1 + b (\alpha_1 + \alpha_2 - \alpha_3))\big)}{\gamma\big(\frac{1}{2} (1 + b^2)\big) \gamma\big(1 + b \alpha_1\big)^2} \cdot \nonumber \\ 
    & \cdot \frac{\gamma\big( \frac{1}{2} b (-b + \alpha_1 + \alpha_2 + \alpha_3)\big) \gamma\big(\frac{1}{2} (1 + b (\alpha_1 - \alpha_2 + \alpha_3))\big)}{\gamma\big(\frac{1}{2} - b^2/2 + b \alpha_1\big) \gamma\big(\frac{1}{2} b (-b - \alpha_1 + \alpha_2 + \alpha_3)\big)} . \label{65}
\end{align}

A similar procedure is valid for the basis correlation function of the odd subsystem. The results presented above can also be applied to $\langle \Psi_{-b}(w, \Bar{w}) \Phi_1(0) \Phi_2(1) \Phi_3(\infty) \rangle$. In this case, the OPE have the field $\Psi$
\begin{align}
    \Psi_{-b} (w, \Bar{w}) \Phi_{\alpha_1} (0) = \sum_s C_{-b, \alpha_1}^{s} |w|^{2(\Delta_s - \Delta_{-b} - \Delta_{\alpha_1})} ( O_{s} (0) + ...) .
\end{align}
Now we have possible classes of fields: two fermionic and one bosonic.
\begin{align} \label{66}
    [\Psi_{-b}]  [\Phi_{\alpha_1}] = [\Psi_{\alpha_1-b}] + [\Phi_{\alpha_1}] + [\Psi_{\alpha_1+b}] .
\end{align}
The correspondence between conformal families and integral representations is similar to the ones above 
\begin{align*}
    [\Psi_{\alpha_1-b}] \rightarrow I_1(w), \hspace{20pt} [\Phi_{\alpha_1}] \rightarrow I_2(w), \hspace{20pt} [\Psi_{\alpha_1+b}] \rightarrow I_3(w) .
\end{align*}
The expressions for the ratio of the structure constants for $\langle \Psi_{-b} \Phi_1 \Phi_2 \Phi_3 \rangle$ have the form:
\begin{align}
    \frac{\widetilde{C}_{-b, \alpha_1}^{\alpha_1 - b} \widetilde{C}(\alpha_1 - b,\alpha_2,\alpha_3)}{\widetilde{C}_{-b, \alpha_1}^{\alpha_1 + b} \widetilde{C}(\alpha_1 + b,\alpha_2,\alpha_3)} = & \frac{1}{4} (1 + b^2 - 2 b \alpha_1)^2 (-1 + b \alpha_1)^2 \frac{ \gamma \big(\frac{1}{2} (1 + b^2 + b (\alpha_1 - \alpha_2 -\alpha_3)) \big) }{ \gamma \big(b \alpha_1 \big) \gamma \big(1 - b^2 + b \alpha_1 \big) } \cdot \nonumber \\
    &\cdot  \frac{\gamma \big( \frac{1}{2} (1 - b^2 + b (\alpha_1 + \alpha_2 - \alpha_3)) \big) \gamma \big(-\frac{1}{2} b (2 b - \alpha_1 - \alpha_2 - \alpha_3) \big) \gamma \big( \frac{1}{2} b (\alpha_1 + \alpha_2 - \alpha_3) \big)}{\gamma \big( \frac{1}{2} - \frac{1}{2} b^2 + b \alpha_1 \big)^2} \cdot \nonumber\\
    &\cdot  \frac{ \gamma \big(1 - \frac{1}{2} b (-\alpha_1 + \alpha_2 + \alpha_3) \big) \gamma \big( \frac{1}{2} (-1 - b^2 + b (\alpha_1 + \alpha_2 + \alpha_3)) \big)}{ \gamma \big(\frac{1}{2} (1 + b^2 - b (\alpha_1 - \alpha_2 + \alpha_3)) \big) \gamma \big(1 - \frac{1}{2} b (\alpha_1 - \alpha_2 + \alpha_3) \big)} , \label{68} \\
    \frac{ C_{-b, \alpha_1}^{\alpha_1} C(\alpha_1, \alpha_2,\alpha_3) }{\widetilde{C}_{-b, \alpha_1}^{\alpha_1 + b} \widetilde{C}(\alpha_1 + b,\alpha_2,\alpha_3)} = & - \frac{(1 + b^2 - 2 b \alpha_1)^2 (-1 + b \alpha_1)^2 }{b^2 (b - \alpha_1)^2 } \frac{\gamma \big(-1 + b^2 \big) \gamma \big(1 - b^2 + b \alpha_1 \big) \gamma \big( \frac{1}{2} b (\alpha_1 + \alpha_2 - \alpha_3) \big)}{ \gamma \big(\frac{1}{2} (-1 + b^2) \big)  \gamma \big(b \alpha_1 \big)^2 } \cdot \nonumber \\
    &\cdot \frac{ \gamma \big( \frac{1}{2} (1 + b^2 + b (\alpha_1 - \alpha_2 - \alpha_3)) \big)  \gamma \big( \frac{1}{2} (-1 - b^2 + b (\alpha_1 + \alpha_2 + \alpha_3)) \big)}{ \gamma \big(\frac{1}{2} - \frac{1}{2} b^2 + b \alpha_1 \big) \gamma \big( 1 - \frac{1}{2} b (\alpha_1 - \alpha_2 + \alpha_3) \big)} . \label{69} 
\end{align}

\section{General expression for three-point correlation function}
Liouville Field Theory is an interacting theory for $\mu \neq 0$. However, as Goulian and Lee \cite{goulian1991correlation} noted, it can be studied by free field methods. In particular, any multipoint correlation function has a pole when the screening condition $\sum \alpha_i + n b = Q$ is satisfied. Moreover, a residue is expressed in terms of the n-dimensional Coulomb integral. This approach can also be applied in SLFT. For case of three-point functions $C(\alpha_1,\alpha_2,\alpha_3)$ and $\widetilde{C} (\alpha_1,\alpha_2,\alpha_3) $, it has the following form
\begin{align} \label{673}
    \textrm{Res} \langle \mathcal{O}_{\alpha_1}(0) \Phi_{\alpha_2}(1) \Phi_{\alpha_3}\infty) \rangle \bigg|_{\sum \alpha_i + n b = Q } = \frac{(-2i \mu b^2)^n}{n!} \mathcal{G}_r(0, 1, \infty) ,
\end{align}
where
\begin{align}
    \mathcal{G}_n (0, 1, \infty) = \int \langle \psi(z_1) \bar{\psi}(z_1) ... \psi(z_m)\bar{\psi}(z_m) \rangle \prod \limits_{k} |z_k|^{-2 \alpha_1 b} |z_k-1|^{-2 \alpha_2 b} \prod \limits_{i < j} |z_i - z_j|^{-2 b^2} d^2 z_k ,
\end{align}
and
\begin{align*}
    m &= 2n, \hspace{20pt} \text{ if } \mathcal{O}_{\alpha_1} = \Phi_{\alpha_1} , \\
    m &= 2n + 1, \text{ if } \mathcal{O}_{\alpha_1} = \Psi_{\alpha_1} .
\end{align*}

At the same time, $C(\alpha_1,\alpha_2,\alpha_3)$ and $\widetilde{C} (\alpha_1,\alpha_2,\alpha_3) $ can be found with help of functional relations (\ref{64}), (\ref{65}) or (\ref{68}), (\ref{69}). There are similar ones, related to the vector (\ref{22}) and the corresponding differential equation. Such dual relation can be obtained after substitution $b \leftrightarrow b^{-1}$. Therefore, the solution can be expressed via the function $\Upsilon(x)$, which is also self dual \cite{zamolodchikov1996conformal}. $\Upsilon(x)$ is defined by the integral representation
\begin{align}
    \log \Upsilon(x) = \int\limits_{0}^{+\infty} \frac{dt}{t} \Bigg( \bigg(\frac{Q}{2}-x \bigg)^2 e^{-t} - \frac{\sinh^2 \big(\frac{Q}{2}-x \big)\frac{t}{2}}{\sinh \frac{bt}{2} \sinh \frac{t}{2b}} \Bigg) ,
\end{align}
and has the following properties
\begin{align}
\begin{split}
    \Upsilon(x + b) &= b^{1- 2 b x} \gamma (b x) \Upsilon(x) , \\
    \Upsilon(Q + \frac{1}{b}) &= b^{\frac{2x}{b}-1} \gamma (x/b) \Upsilon(x) , \\
    \Upsilon(Q - x) &= \Upsilon(x) , \\
    \Upsilon(Q/2) &= 1 . \\
\end{split}
\end{align}
Moreover, $\Upsilon(x)$ has zeroes that located at $x = - mb^{-1} - n b$ and $x = (m + 1)b^{-1} - (n + 1) b$, where m and n run over all non-negative integers. 

Let us focus our attention on the functional relations associated with the even subsystem. Firstly, let us consider (\ref{64}) and find $C(\alpha_1,\alpha_2,\alpha_3)$. It is easy to see that the solution of (\ref{64}) should be expressed through $\Upsilon(x)$, or more precisely their combination, up to the product of normalization factors. General expression has form
\begin{align} \label{83}
    &C(\alpha_1,\alpha_2,\alpha_3) = \nonumber \\
    &\frac{e^{\lambda (Q - \sum \alpha_i )} \mathcal{N}(\alpha_1) \mathcal{N}(\alpha_2) \mathcal{N}(\alpha_3)}{ \underset{j,k \neq i}{\underset{i=1}{\prod \limits^3}} \bigg( \Upsilon \big(\frac{1}{2} (-\alpha_i + \alpha_j + \alpha_k) \big) \Upsilon \big(\frac{1}{2} (Q -\alpha_i + \alpha_j + \alpha_k) \big) \bigg) \Upsilon \big(
   \frac{1}{2} (\alpha_1 + \alpha_2 + \alpha_3) \big) \Upsilon \big(
   \frac{1}{2} (-Q + \alpha_1 + \alpha_2 + \alpha_3) \big) } .
\end{align}
In turn, normalization constants can be found using the free field method. It is enough to use (\ref{673}) for $n=0$ and $n=2$
\begin{align} 
    \textrm{Res} \langle& \Phi_{\alpha_1}(0) \Phi_{\alpha_2}(1) \Phi_{\alpha_3}\infty) \rangle \bigg|_{\sum \alpha_i = Q } = 1 , \label{76} \\
    \textrm{Res} \langle \Phi_{\alpha_1}(0) \Phi_{\alpha_2}(1) \Phi_{\alpha_3}\infty) \rangle \bigg|_{\sum \alpha_i + 2b = Q } &= 2 \mu^2 \pi^2 b^4 \frac{ \gamma (-b Q)}{ \gamma (-\frac{1}{2} b Q )} \prod \limits_{i=1}^3 \gamma \big(b (Q - \alpha_i - b) \big) \gamma \big(b (Q/2 - \alpha_i - b) \big) . \label{77}
\end{align}
It is important to note that the calculation with the charge balance $Q= \sum \alpha_i$ does not allow us to fix the normalization factor completely.  It can also be noted that the normalization should contain a factor associated with the cosmological constant $\mu$. These arguments explain the specific form of the additional normalization that has an exponential form in (\ref{83}). The first calculation with $Q= \sum \alpha_i$ allows to fix $\mathcal{N}(\alpha_k)$ up to an exponential factor $e^{\lambda\left(\frac{Q}{3}-\alpha_k\right)}$, since
\begin{align*}
    \prod_{k=1}^3e^{\lambda\left(\frac{Q}{3}-\alpha_k\right)}\Bigl|_{\sum\alpha_i=Q}=1,
\end{align*}
for any $\lambda$. The second calculation with $Q= \sum \alpha_i + 2b$ determines $\lambda (\mu)$. Using (\ref{83}), (\ref{76}) and (\ref{77}) we find that unknown normalization factor obey the relations
\begin{align}
    &\mathcal{N}(\alpha_1) \mathcal{N}(\alpha_2) \mathcal{N}(\alpha_3) = \Upsilon_{0} \prod \limits_{i=1}^3 \Upsilon \bigg( \frac{1}{2} Q - \alpha_i \bigg) \Upsilon ( \alpha_i ) , \\
    &\lambda (\mu) = \ln \bigg( \frac{ 1}{\sqrt{2}} \pi \mu b^{1 - b^2} \gamma \bigg( \frac{1}{2} + \frac{1}{2} b^2 \bigg) \bigg)^{\frac{1}{b}} ,
\end{align}
where the following notation is introduced
\begin{align*}
    \Upsilon_{0} = \frac{d \Upsilon(x)}{dx} \bigg|_{x = 0} .
\end{align*}

Thus, the general form of $C(\alpha_1,\alpha_2,\alpha_3)$ is
\begin{align}
    &C(\alpha_1,\alpha_2,\alpha_3) = \bigg( \frac{ 1}{\sqrt{2}} \pi \mu b^{1 - b^2} \gamma \bigg( \frac{1}{2} + \frac{1}{2} b^2 \bigg) \bigg)^{\frac{Q - \sum \alpha_i }{b}} \cdot \nonumber \\
    &\cdot \frac{\Upsilon_{0} \prod \limits_{i=1}^3 \bigg( \Upsilon(\frac{1}{2}Q - \alpha_i) \Upsilon(\alpha_i) \bigg)}{ \underset{j,k \neq i}{\underset{i=1}{\prod \limits^3}} \bigg( \Upsilon \big(\frac{1}{2} (-\alpha_i + \alpha_j + \alpha_k) \big) \Upsilon \big(\frac{1}{2} (Q -\alpha_i + \alpha_j + \alpha_k) \big) \bigg) \Upsilon \big(
   \frac{1}{2} (\alpha_1 + \alpha_2 + \alpha_3) \big) \Upsilon \big(
   \frac{1}{2} (-Q + \alpha_1 + \alpha_2 + \alpha_3) \big) }.\label{C-SLFT-DOZZ}
\end{align}
It can be shown that \eqref{C-SLFT-DOZZ} satisfies \eqref{673} for any $n$ (see \cite{rashkov1996three} for details).

Now, let us move on to the next part related to (\ref{65}) and $\widetilde{C}(\alpha_1, \alpha_2,\alpha_3)$.
It is easy to note that the correlation function differs from the boson function only by a normalization. Moreover, this factor depends on one parameter $\alpha_1$. In this case, it is sufficient to consider the screening condition $\sum \alpha_i + b = Q$, since the dependence on $\mu$ is known and does not change fundamentally
\begin{align}
    \textrm{Res} \langle \Psi_{\alpha_1}(0) \Phi_{\alpha_2}(1) \Phi_{\alpha_3}\infty) \rangle \bigg|_{\sum \alpha_i +b = Q } = -2i \mu \pi b^2 \alpha_1^2 \frac{\gamma \big( b (Q-\alpha_2-b)  \big) \gamma \big( b (Q-\alpha_3-b)  \big)}{\gamma \big( b (Q+\alpha_1-b)  \big)} .
\end{align}
After simplification, expression for three-point correlation function can be received
\begin{align}
    &\widetilde{C}(\alpha_1, \alpha_2,\alpha_3) = -2 \sqrt{2} i \alpha_1^2 \frac{\Upsilon(-\alpha_1 + b)}{\Upsilon(-\alpha_1)} \bigg( \frac{ 1}{\sqrt{2}} \pi \mu b^{1 - b^2} \gamma \bigg( \frac{1}{2} + \frac{1}{2} b^2 \bigg) \bigg)^{\frac{Q - \sum \alpha_i }{b}} \frac{1}{ \Upsilon \big(
   \frac{1}{2} (\alpha_1 + \alpha_2 + \alpha_3 - b) \big) }\cdot \nonumber \\
    &\cdot \frac{\Upsilon_{0} \Upsilon(\frac{1}{2}Q - \alpha_1 ) \Upsilon( \alpha_1 + b)\prod \limits_{i=2}^3 \bigg( \Upsilon(\frac{1}{2}Q - \alpha_i) \Upsilon(\alpha_i) \bigg)}{ \Upsilon \big(
   \frac{1}{2} (-Q + \alpha_1 + \alpha_2 + \alpha_3 + b) \big) \underset{j,k \neq i}{\underset{i=1}{\prod \limits^3}} \bigg( \Upsilon \big(\frac{1}{2} (-\alpha_i + \alpha_j + \alpha_k + b) \big) \Upsilon \big(\frac{1}{2} (Q -\alpha_i + \alpha_j + \alpha_k - b) \big) \bigg) } .
\end{align}
Similar expressions were found in \cite{poghossian1997structure}, but using Ramond fields. Probably, the differences are due to a different normalization. As a result, three-point correlation functions are presented. Others can be found in a similar way, described in \hyperref[sec: Chapter 3]{section 3} and related to the Ward identities

\section*{Conclusion}
In this paper the correlation functions involving degenerate fields in $NS$ sector were studied. Four-point correlation functions with one degenerate field at level $3/2$ were shown to have integral form, corresponding to the special Dotsenko-Fateev equation. 

Very interestingly, for a vector at the level $N= \frac{3}{2}$ there is a coincidence with the boson theory and its methods. Thus, integral solutions can be represented as conformal blocks using the method of reducing the number of integrations \cite{fateev2008multipoint}. In the future, we would like to study the differential equations obtained from the vectors at the next levels and their solutions. This will allow us to understand the more general structure.

In addition, the general properties of operator expansions were studied and verified using integral solutions as an example. Thus, the relations for the structural constants made it possible to obtain general expressions for the basic correlation functions of three-point functions.

\section*{Acknowledgments}
I would like to thank Alexey Litvinov for useful collaboration and valuable discussions.

\section*{Apendix A} \label{sec: Apendix A}
In this appendix we present explicit expressions for four-point correlation functions of odd and even subsystems studied in \hyperref[sec: Chapter 3]{section 3}. In the formulas below, the indices are omitted for ease of notation.

\subsection*{Even subsystem}
Firsly, we represent even correlation functions
\begin{align}
    \langle \Psi \Psi \Phi \Phi \rangle =& \frac{1}{F_{01}} \bigg(-A \partial^2  + B(x_1, x_2, x_3) \partial + 2b^{2} C_1 (x_1, \Delta_1, x_2, \Delta_2, x_3, \Delta_3) \bigg) \langle \Phi \Phi \Phi \Phi \rangle , \\
    \langle \Psi \Phi \Psi \Phi \rangle =& \frac{1}{F_{02}} \bigg(-A \partial^2  + B(x_2, x_3, x_1) \partial + 2b^{2}  C_1 ( x_2, \Delta_2, x_3, \Delta_3, x_1, \Delta_1) \bigg) \langle \Phi \Phi \Phi \Phi \rangle , \\
    \langle \Psi \Phi \Phi \Psi \rangle =& \frac{1}{F_{03}} \bigg(-A \partial^2  + B(x_3, x_1, x_2) \partial + 2b^{2}  C_1 (x_3, \Delta_3, x_1, \Delta_1, x_2, \Delta_2) \bigg) \langle \Phi \Phi \Phi \Phi \rangle , \\
    \langle \Phi \Psi \Psi \Phi \rangle =& \frac{1}{F_{13}} \bigg(-A \partial^2  + B(x_3, x_1, x_2) \partial + b^{2}  C_2 (x_3, \Delta_3, x_1, \Delta_1, x_2, \Delta_2) \bigg) \langle \Phi \Phi \Phi \Phi \rangle , \\
    \langle \Phi \Psi \Phi \Psi \rangle =& \frac{1}{F_{13}} \bigg(-A \partial^2  + B(x_2, x_3, x_1) \partial + b^{2} C_2 (x_2, \Delta_2, x_3, \Delta_3, x_1, \Delta_1) \bigg) \langle \Phi \Phi \Phi \Phi \rangle , \\
    \langle \Phi \Phi \Psi \Psi \rangle =& \frac{1}{F_{23}} \bigg(-A \partial^2  + B(x_1, x_2, x_3) \partial + b^{2} C_2 (x_1, \Delta_1, x_2, \Delta_2, x_3, \Delta_3)  \bigg) \langle \Phi \Phi \Phi \Phi \rangle ,
\end{align}

where
\begin{align}
    &A = (x - x_1)^2(x - x_2)^2(x - x_3)^2 , \\
    &B (x_1, x_2, x_3)= b^2 (x - x_1) (x - x_2) (x - x_3) \cdot \nonumber\\ 
    &\hspace{2,3cm} \cdot \big(4 x^2 - 2 x x_1 + 2 x_2 x_3 + (- 3 x + x_1) (x_2 + x_3) \big) , \\
    &C_1 (x_1, \Delta_1, x_2, \Delta_2, x_3, \Delta_3) = 2 x^4 \Delta - x^3 \big(2 x_1 + 3 (x_2 + x_3) \big) \Delta + \nonumber \\
    &\hspace{2,3cm} + x^2 \big( -x_1^2 \Delta_1 + 
   x_2^2 (\Delta - \Delta_2) + 
   x_3^2 (\Delta - \Delta_3) + \nonumber \\
   &\hspace{2,3cm} + x_1 x_2 (3 \Delta + \Delta_1 + \Delta_2 - \Delta_3) + x_1 x_3 (3 \Delta + \Delta_1 - \Delta_2 + \Delta_3) + \nonumber \\ 
   &\hspace{2,3cm} + x_2 x_3 (4 \Delta - \Delta_1 + \Delta_2 + \Delta_3) \big) + \nonumber \\
   &\hspace{2,3cm} + x \bigg( x_1^2 \big(x_3 (\Delta_1 + \Delta_2 - \Delta_3) + x_2 (\Delta_1 - \Delta_2 + \Delta_3) \big) + \nonumber \\ 
   &\hspace{2,3cm} +x_2 x_3 \big(x_2 (-\Delta + \Delta_1 + \Delta_2 - \Delta_3) + x_3 (-\Delta + \Delta_1 - \Delta_2 + \Delta_3) \big) - \nonumber \\
   &\hspace{2,3cm} -x_1 \big((x_2^2 + x_3^2) (\Delta + \Delta_1 -\Delta_2 - \Delta_3) + 2 x_2 x_3 (2 \Delta + \Delta_1 + \Delta_2 + \Delta_3) \big) \bigg) \nonumber \\
    &\hspace{2,3cm} -(x_1^2 x_2^2 \Delta_3 + x_1^2 x_3^2 \Delta_2 + x_2^2 x_3^2 \Delta_1) + x_1 x_2 x_3 \big(x_1 (-\Delta_1 + \Delta_2 + \Delta_3) + \nonumber \\
    &\hspace{2,3cm} + x_2 (\Delta + \Delta_1 - \Delta_2 + \Delta_3) + x_3 (\Delta + \Delta_1 + \Delta_2 - \Delta_3) \big) , \\
    &C_2 (x_1, \Delta_1, x_2, \Delta_2, x_3, \Delta_3) = 4 x^4 \Delta - 2 x^3 \big(2 x_1 + 3 (x_2 + x_3) \big) \Delta + \nonumber \\
    &\hspace{2,3cm} + x^2 \bigg(2 x_2^2 (\Delta - \Delta_2) + 2 x_3^2 (\Delta - \Delta_3) + x_1^2 (\Delta - \Delta_1 - \Delta_2 - \Delta_3) + \nonumber \\ 
    &\hspace{2,3cm} + x_1 \big(x_2 (5 \Delta + \Delta_1 + 3 \Delta_2 - \Delta_3) + x_3 (5 \Delta + \Delta_1 - \Delta_2 + 3 \Delta_3) \big) + \nonumber \\
     &\hspace{2,3cm} + x_2 x_3 (9 \Delta - \Delta_1 + \Delta_2 + \Delta_3) \bigg) + \nonumber \\
    &\hspace{2,3cm} + x \bigg(x_1^2 \big(x_2 (\Delta - \Delta_1 + \Delta_2 - 3 \Delta_3) + x_3 (\Delta - \Delta_1 - 3 \Delta_2 + \Delta_3) \big) + \nonumber \\
    &\hspace{2,3cm} + x_2 x_3 \big(x_3 (3 \Delta - \Delta_1 + \Delta_2 - 3 \Delta_3) + x_2 (3 \Delta - \Delta_1 - 3 \Delta_2 + \Delta_3) \big) + \nonumber \\ 
    &\hspace{2,3cm} + x_1 \big((x_2^2 + x_3^2) (\Delta + \Delta_1 - \Delta_2 - \Delta_3) + 2 x_2 x_3 (3 \Delta + \Delta_1 + 3 (\Delta_2 + \Delta_3))\big) \bigg) \nonumber \\
    &\hspace{2,3cm} - 2 (x_2^2 x_3^2 \Delta_1 + x_1^2 x_3^2 \Delta_2 + x_1^2 x_2^2 \Delta_3) + \nonumber \\ 
    &\hspace{2,3cm} + (x_1^2 x_2 x_3 (\Delta - \Delta_1 + \Delta_2 + \Delta_3) + x_2^2 x_3^2 (\Delta + \Delta_1 - \Delta_2 - \Delta_3)) + \nonumber \\ 
    &\hspace{2,3cm} + x_1 x_2 x_3 ( x_3 (\Delta + \Delta_1 + 3 \Delta_2 - \Delta_3) + x_2 (\Delta + \Delta_1 - \Delta_2 + 3 \Delta_3)) , \\
    &F_{ij}= b^2 \underset{l \neq i}{\underset{l=0}{\prod^3}}(x_i - x_l) \underset{l \neq i,j}{\underset{l=0}{\prod^3}}(x_j - x_l) .
\end{align}
The expression for the fermion function is calculated separately.:
\begin{align}
      \langle \Psi \Psi \Psi \Psi \rangle = \frac{1}{b^2 \underset{l,i \in \overline{0,3}}{\underset{l < i}{\prod}}(x_l - x_l)} (A_3 \partial^2 - 2b^2 B_3 \partial - 2b^2 C_3) \langle \Phi \Phi \Phi \Phi \rangle,
\end{align}
where
\begin{align}
    A_3 =& (x - x_1)^2 (x - x_2)^2 (x - 
    x_3)^2 \bigg(-1 + \Delta + \sum_{i = 1}^3 \Delta_i + b^2 \bigg) , \\
    B_3 =& (x - x_1) (x - x_2) (x - 
    x_3) \Bigg( \sum_{i < j} x_i x_j (-1 + \Delta_i + \Delta_j ) + x^2 \bigg( -3 + 2 \sum_{i = 1}^3 \Delta_i \bigg) - \nonumber \\
    & - x \sum_{i=1}^3 x_i (-2 + 2 \Delta_i + \Delta_k + \Delta_l )  \Bigg) , \\
    C_3 =& x^4 \Delta \bigg( -3 + 2 \sum_{i = 1}^3 \Delta_i \bigg) -  2 x^3 \Delta \sum_i x_i (-2 + 2 \Delta_i + \Delta_k + \Delta_l ) + \nonumber \\
    &+ x^2 \Bigg( \sum_{i=1}^3 \bigg( 
     x_i^2 \big(\Delta (-1 + \Delta_i) - \Delta_i (-1 + \sum_{i = 1}^3 \Delta_i) \big) \bigg) + \nonumber \\
     & +\sum_{i<j; m \neq i,j} x_i x_j \bigg(\Delta_m (1 - \Delta_m) + (\Delta_i + \Delta_j - 1) (\Delta_i + \Delta_j) + \nonumber
\end{align}
\begin{align}
    \hspace{-1.1cm} &+\Delta \big(\Delta_m + 5 (-1 + \Delta_i + \Delta_j) \big) \bigg) \Bigg) + x \Bigg(\sum_{i=1}^3 \bigg(x_i^2 x_j \big(\Delta_j (1 - \Delta_j) + \nonumber \\
    \hspace{-1.1cm} &+ (\Delta_i + \Delta_m - 1) (\Delta_i + \Delta_m) + \Delta (1 + \Delta_m - \Delta_i - \Delta_j) \big) \bigg) - \nonumber \\
    \hspace{-1.1cm} &- 2 x_1 x_2 x_3 (-1 + \sum_{i = 1}^3 \Delta_i) (3 \Delta + \sum_{i = 1}^3 \Delta_i) \Bigg) - \nonumber \\
    \hspace{-1.1cm} &- \bigg( -1 + \Delta + \sum_{i = 1}^3 \Delta_i \bigg) \bigg(\sum_{i,j,k \in \sigma(1,2,3)} x_i^2 x_j^2 \Delta_k \bigg) + \nonumber \\
    \hspace{-1.1cm} &+ \bigg(-1 + \sum_{i = 1}^3 \Delta_i \bigg) \bigg( \sum x_i^2 x_k x_l (\Delta - \Delta_i + \Delta_k + \Delta_l ) \bigg) ,
\end{align}
where $\sigma(s_1,s_2,...)$ -- cyclic permutation group; the set of indices $k,l$ satisfies the following conditions: \{$k,l \neq i$;  $k<l$; $k,l \in \overline{1,3} $\}

\subsection*{Odd subsystem}
Secondly, we represent odd correlation functions
\begin{align}
    \langle \Phi \Psi \Phi \Phi \rangle =& - F_1 (x_1 | x_2, x_3) \bigg( A_1 (x_1 | x_2, x_3) \partial + B_1 (x_1 | x_2, x_3) \bigg) \langle \Psi \Phi \Phi \Phi \rangle , \hspace{2,2cm} \\
    \langle \Phi \Phi \Psi \Phi \rangle =& - F_1 (x_2 | x_1, x_3) \bigg( A_1 (x_2 | x_1, x_3) \partial + B_1 (x_2 | x_1, x_3) \bigg) \langle \Psi \Phi \Phi \Phi \rangle , 
\end{align}
\begin{align}
    \langle \Phi \Phi \Phi \Psi \rangle =& - F_1 (x_3 | x_1, x_2) \bigg( A_1 (x_3 | x_1, x_2) \partial + B_1 (x_3 | x_1, x_2) \bigg) \langle \Psi \Phi \Phi \Phi \rangle , \\
    \langle \Psi \Phi \Psi \Psi \rangle =& F_2 \bigg( A_2 (x_1 | x_2, x_3) \partial^2 + B_2 (x_1 | x_2, x_3) \partial + C_2 (x_1 | x_2, x_3) \bigg) \langle \Psi \Phi \Phi \Phi \rangle , \hspace{1,4cm} \\
    \langle \Psi \Psi \Phi \Psi \rangle =& - F_2 \bigg( A_2 (x_2 | x_1, x_3) \partial^2 + B_2 (x_2 | x_1, x_3) \partial + C_2 (x_2 | x_1, x_3) \bigg) \langle \Psi \Phi \Phi \Phi \rangle , \\
    \langle \Psi \Psi \Psi \Phi \rangle =& F_2 \bigg( A_2 (x_3 | x_1, x_2) \partial^2 + B_2 (x_3 | x_1, x_2) \partial + C_2 (x_3 | x_1, x_2) \bigg) \langle \Psi \Phi \Phi \Phi \rangle , \\
    \langle \Phi \Psi \Psi \Psi \rangle =& \frac{1}{2 b^2 \underset{i,j \in \overline{1,3}}{\underset{i < j}{\prod}}(x_i - x_j) } \Bigg( \prod_{i=1}^3 (x - x_i) \bigg(1 + 2b^2 + 2 \Delta - 2 \sum_{i=1}^3 \Delta_i \bigg) \partial  + \nonumber \\
     4b^2& \bigg( \underset{i,j,k \in \overline{1,3}}{\underset{i < j; \ k \neq i,j}{\sum}} x_i x_j \Delta_k + x^2 \sum_{i=1}^3 \Delta_i - x \big( \sum_{i,j,k \in \sigma(1,2,3)} x_i (\Delta_j + \Delta_k) \big) \bigg) \Bigg) \langle \Psi \Phi \Phi \Phi \rangle ,
\end{align}
where $\sigma(s_1,s_2,...)$ -- cyclic permutation group, and coefficients have form:
\begin{align}
    &A_1 (x_1 | x_2, x_3) = (x - x_2)(x - x_3) , \\
    &B_1 (x_1 | x_2, x_3) = b^2 (-2x + x_2 + x_3) , \\
    &F_1 (x_1 | x_2, x_3) = \frac{x - x_1}{b^2 (x_1 - x_2)(x_1 - x_3)} , \\
    &A_2 (x_1 | x_2, x_3) =  2 (x - x_1) \prod_{i=1}^3 (x - x_i) , \\
    &B_2 (x_1 | x_2, x_3) = 2 (x - x_1) \bigg( x^2 (3 + 2 \Delta) - 2 x \big(x_1 + (x_2 + x_3) (1 + \Delta) \big) + \nonumber \\
    &\hspace{2,5cm} +x_1 (x_2 + x_3) + 
   x_2 x_3 (1 + 2 \Delta) - b^2 (x - x_1) (2 x - x_2 - x_3) \bigg) , \\
    &C_2 (x_1 | x_2, x_3) = b^2 \bigg(-8 x^2 (1 + \Delta) + 2 x \big(4 x_1 + (2 x_1 + x_2 + x_3) (1 + 2 \Delta) \big) + \nonumber \\
    &\hspace{2,5cm} + x_2 x_3 (1 - 2 \Delta + 2 \Delta_1 - 2 \Delta_2 - 2 \Delta_3) - \nonumber \\
    &\hspace{2,5cm} - x_1 (x_2 + x_3) (3 + 2 \Delta + 2 \Delta_1 - 2 \Delta_2 - 2 \Delta_3) - \nonumber \\
    &\hspace{2,5cm} - x_1^2 (3 + 2 \Delta - 2 \Delta_1 + 2 \Delta_2 + 2 \Delta_3) \bigg) , \\
    &F_2 = \frac{1}{2 b^2 \underset{i,j \in \overline{1,3}}{\underset{i < j}{\prod}}(x_i - x_j) } . \hspace{9cm}
\end{align}

\section*{Apendix B} \label{sec: Apendix B}
As it was mentioned in \hyperref[sec: Chapter 4]{section 4}, the hypergeometric substitution allows us to reduce the equation to the following form \cite{dotsenko1984conformal}
\begin{align} \label{127}
\begin{split}
    &x^2 (x-1)^2 F'''(x) + \big( K_1 x + K_2 (x-1) \big) x (x-1) F''(x) + \\
    &+\big( L_1 x^2 + L_2 (x - 1)^2 + L_3 x(x-1) \big) F'(x) + (M_1 x + M_2 (x-1)) F(x) = 0,
\end{split}
\end{align}
where
\begin{align*}
    K_1 =& - (g + 3 B + 3 C), \hspace{70pt}
    K_2 = -(g + 3 A + 3 c), \\
    L_1 =& (B + C) (2 B + 2 C + g + 1), \hspace{23pt}
    L_2 = (A + C) (2 A + 2 C + g + 1), \\
    L_3 =& (B + C) (2 A + 2 C + g + 1) + (A + C) (2 B + 2 C + g + 1) + \\
    &+(C - 1) (A + B + C) + (3 C + g) (A + B + C + g + 1), \\
    M_1 =& -C (2 B + 2 C + g + 1) (2 A + 2 B + 2 C + g + 2), \\
    M_2 =& -C (2 A + 2 C + g + 1) (2 A + 2 B + 2 C + g + 2),
\end{align*}
The solution (\ref{127}) can be generally represented as a double integral with a set of parameters $ \{ A,B,C,g \}$
\begin{align} \label{(4.5)}
    I(A,B,C,g;z) = \int\limits_{C1} dt_1 \int\limits_{C2} dt_2 t_1^A (t_1 - 1)^B (t_1 - z)^C  t_2^A (t_2 - 1)^B (t_2 - z)^C (t_1 - t_2)^g .
\end{align}

The physical correlation function have form
\begin{align} \label{128}
    G(w, \Bar{w}) = \sum X_{i} I_i(w) \overline{I_j(w)} .
\end{align}
$I_i(w)$ has three singularities $0, \ 1, \ \infty$. If closed contour surrounds this point, the integral transfor under the action of the monodromy group. It is clear that the physical correlation functions $G(w, \Bar{w})$ must be invariant. The required invariance imposes condition of diagonality of the transformation matrix. Also, there are three independent configurations of contours
\begin{align}
\begin{split}
    &I_1(w)  \longmapsto C_1 = [1,+\infty) , C_2 =[1,+\infty) ; \\
    &I_2(w)  \longmapsto C_1 = [1,+\infty) , C_2 = [0,w] ;\\
    &I_3(w)  \longmapsto C_1 = [0,w] , C_2 = [0,w] .
\end{split}
\end{align}
The function $G(w, \Bar{w})$ has the form \begin{align} \label{(4.11)}
    G(w, \Bar{w}) = X_{1} |I_1(w)|^2 + X_{2} |I_2(w)|^2 + X_{3} |I_3(w)|^2, 
\end{align}

To find the $X_i/X_j$, it is necessary to expand the basis $I_i(w)$ at $z=0$ via the basis $\widetilde{I_i}(1-w)$ at $z=1$:
\begin{align} \label{131}
    I_i(w) = \sum_j \alpha_{ij} I_j(1-w)
\end{align}
Substituting the expansion (\ref{131}) in (\ref{128}) and the diagonality of the monodromy matrix lead to next conditions
\begin{align} \label{(4.16)}
    \sum X_{i} \alpha_{ik} \alpha_{il} = 0, \hspace{20pt} l \neq k .
\end{align}
Manipulation with the contours allows to determine constants $\alpha_{ij}$
\begin{align*}
    &\alpha_{11} = \frac{\sin{\pi (A) } \sin{\pi (A + \frac{1}{2}g) } }{\sin{\pi (B+C) } \sin{\pi (B+C+\frac{1}{2}g) } } ,  \\
    &\alpha_{12} = \frac{\sin{\pi (A) } \sin{\pi (C) } }{\sin{\pi (B+C) } \sin{\pi (B+C+g) } } ,  \\
    &\alpha_{13} = \frac{\sin{\pi (C) } \sin{\pi (C+\frac{1}{2}g) } }{\sin{\pi (B+C+\frac{1}{2}g) } \sin{\pi (B+C+g) } }  , \\
    &\alpha_{21} = - \frac{\sin{\pi (A+B+C+\frac{1}{2}g) } \sin{\pi (A+\frac{1}{2}g) 2 \cos{\pi \frac{1}{2}g} } }{\sin{\pi (B+C) } \sin{\pi (B+C+\frac{1}{2}g) } } ,  \\
    &\alpha_{22} = \frac{\sin{\pi (A+B+C+\frac{1}{2}g) } \sin{\pi (C) } }{\sin{\pi (B+C) } \sin{\pi (B+C+\frac{1}{2}g) } } - \frac{\sin{\pi (B+\frac{1}{2}g) } \sin{\pi (A) } }{\sin{\pi (B+C+\frac{1}{2}g) } \sin{\pi (B+C+g) } } , \\
    &\alpha_{23} = \frac{\sin{\pi (B+\frac{1}{2}g) } \sin{\pi (C+\frac{1}{2}g) } 2 \cos{\pi \frac{1}{2}g }}{\sin{\pi (B+C+\frac{1}{2}g) } \sin{\pi (B+C+g) } }  , \\
    &\alpha_{31} = \frac{\sin{\pi (A+B+C+\frac{1}{2}g) } \sin{\pi (A+B+C+g) } }{\sin{\pi (B+C) } \sin{\pi (B+C+\frac{1}{2}g) } } ,  \\
    &\alpha_{32} = - \frac{\sin{\pi (A+B+C+g) } \sin{\pi (B) } }{\sin{\pi (B+C) } \sin{\pi (B+C+g) } } ,  \\
    &\alpha_{33} = \frac{\sin{\pi (B) } \sin{\pi (B+\frac{1}{2}g) } }{\sin{\pi (B+C+\frac{1}{2}g) } \sin{\pi (B+C+g) } }   .
\end{align*}
Finally, constraints fix $X_i/X_j$
\begin{align} \label{(4.18)}
    &\frac{X_1}{X_3} = \frac{\sin{\pi (A+B+C+g) } \sin{\pi (A+B+C+ \frac{1}{2} g) } \sin{\pi (B) } \sin{\pi (B+\frac{1}{2} g) } \sin{\pi (A+C+g) } }{\sin{\pi (A) } \sin{\pi (A+\frac{1}{2} g) } \sin{\pi (C) } \sin{\pi (C+\frac{1}{2} g) } \sin{\pi (A+C) }} , \\
    &\frac{X_2}{X_3} = \frac{\sin{\pi (A+B+C+g) } \sin{\pi (A+C+\frac{1}{2} g) } \sin{\pi (B) } }{\sin{\pi (C+\frac{1}{2} g) } \sin{\pi (A+\frac{1}{2} g) } \sin{\pi (A+C) } 2 \cos{\pi (\frac{1}{2} g) } } . \nonumber
\end{align}
\bibliographystyle{MyStyle}
\bibliography{bibliography}

\providecommand{\href}[2]{#2}\begingroup\raggedright\begin{thebibliography}{10}

\bibitem{polyakov1981quantumb}
A.~M. Polyakov, \emph{Quantum geometry of bosonic strings}, {\emph{Physics Letters B} {\bfseries 103} (1981) 207}.

\bibitem{polyakov1981quantumf}
A.~M. Polyakov, \emph{Quantum geometry of fermionic strings}, {\emph{Physics Letters B} {\bfseries 103} (1981) 211}.

\bibitem{goulian1991correlation}
M.~Goulian and M.~Li, \emph{Correlation functions in liouville theory}, {\emph{Physical Review Letters} {\bfseries 66} (1991) 2051}.

\bibitem{dotsenko1985four}
V.~S. Dotsenko and V.~A. Fateev, \emph{{Four Point Correlation Functions and the Operator Algebra in the Two-Dimensional Conformal Invariant Theories with the Central Charge c \ensuremath{\leq} 1}}, \href{https://doi.org/10.1016/S0550-3213(85)80004-3}{\emph{Nucl. Phys. B} {\bfseries 251} (1985) 691}.

\bibitem{belavin1984infinite}
A.~A. Belavin, A.~M. Polyakov and A.~B. Zamolodchikov, \emph{Infinite conformal symmetry in two-dimensional quantum field theory}, {\emph{Nuclear Physics B} {\bfseries 241} (1984) 333}.

\bibitem{teschner1995liouville}
J.~Teschner, \emph{On the liouville three-point function}, {\emph{Physics Letters B} {\bfseries 363} (1995) 65}.

\bibitem{poghossian1997structure}
R.~H. Poghossian, \emph{{Structure constants in the N=1 superLiouville field theory}}, \href{https://doi.org/10.1016/S0550-3213(97)00218-6}{\emph{Nucl. Phys. B} {\bfseries 496} (1997) 451} [\href{https://arxiv.org/abs/hep-th/9607120}{{\ttfamily hep-th/9607120}}].

\bibitem{friedan1985superconformal}
D.~Friedan, Z.~Qiu and S.~Shenker, \emph{Superconformal invariance in two dimensions and the tricritical ising model}, {\emph{Physics Letters B} {\bfseries 151} (1985) 37}.

\bibitem{mussardo1988fine}
G.~Mussardo, G.~Sotkov and M.~Stanishkov, \emph{Fine structure of the supersymmetric operator product expansion algebras}, {\emph{Nuclear Physics B} {\bfseries 305} (1988) 69}.

\bibitem{fukuda2002super}
T.~Fukuda and K.~Hosomichi, \emph{Super-liouville theory with boundary}, {\emph{Nuclear Physics B} {\bfseries 635} (2002) 215}.

\bibitem{distler1990super}
J.~Distler, Z.~Hlousek and H.~Kawai, \emph{{Superliouville Theory as a Two-Dimensional, Superconformal Supergravity Theory}}, \href{https://doi.org/10.1142/S0217751X90000180}{\emph{Int. J. Mod. Phys. A} {\bfseries 5} (1990) 391}.

\bibitem{neveu1971factorizable}
A.~Neveu and J.~H. Schwarz, \emph{Factorizable dual model of pions}, {\emph{Nuclear Physics B} {\bfseries 31} (1971) 86}.

\bibitem{ramond1971dual}
P.~Ramond, \emph{Dual theory for free fermions}, {\emph{Physical Review D} {\bfseries 3} (1971) 2415}.

\bibitem{dotsenko1984conformal}
V.~S. Dotsenko and V.~A. Fateev, \emph{Conformal algebra and multipoint correlation functions in 2d statistical models}, {\emph{Nuclear Physics B} {\bfseries 240} (1984) 312}.

\bibitem{selberg1944remarks}
A.~Selberg, \emph{Remarks on a multiple integral}, {\emph{Norsk Mat. Tidsskr.} {\bfseries 26} (1944) 71}.

\bibitem{zamolodchikov1996conformal}
A.~Zamolodchikov and A.~Zamolodchikov, \emph{Conformal bootstrap in liouville field theory}, {\emph{Nuclear Physics B} {\bfseries 477} (1996) 577}.

\bibitem{rashkov1996three}
R.~Rashkov and M.~Stanishkov, \emph{Three-point correlation functions in n= 1 super liouville theory}, {\emph{Physics Letters B} {\bfseries 380} (1996) 49}.

\bibitem{fateev2008multipoint}
V.~A. Fateev and A.~V. Litvinov, \emph{{Multipoint correlation functions in Liouville field theory and minimal Liouville gravity}}, \href{https://doi.org/10.1007/s11232-008-0038-3}{\emph{Theor. Math. Phys.} {\bfseries 154} (2008) 454} [\href{https://arxiv.org/abs/0707.1664}{{\ttfamily 0707.1664}}].

\end{thebibliography}\endgroup

\end{document}